\documentclass[prd,preprint,showpacs,preprintnumbers,amsmath,amssymb,eqsecnum]{revtex4-1}

\usepackage{graphicx}
\usepackage{dcolumn}
\usepackage{bm}
\usepackage{subfigure}
\usepackage{color}

\begin{document}

\preprint{RUP-12-4}

\title{Upper limits of particle emission 
from high-energy collision and reaction 
near a maximally rotating Kerr black hole
}
\author{$^{1,2}$Tomohiro Harada}\email{harada@rikkyo.ac.jp}
\author{$^{1}$Hiroya Nemoto} 
\author{$^{1}$Umpei Miyamoto}
\date{\today}
\affiliation{%
$^{1}$Department of Physics, Rikkyo University, Toshima, Tokyo 171-8501, Japan
}%
\affiliation{
$^{2}$Astronomy Unit, School of Physics and Astronomy, Queen Mary, University of London, Mile End Road, London E1 4NS, UK
}
\begin{abstract}
The center-of-mass energy of two particles 
colliding near the horizon of a
maximally rotating black hole can be arbitrarily high if 
the angular momentum of either of the incident 
particles is fine-tuned, 
which we call a critical particle. 
We study particle emission from such high-energy collision
and reaction in the equatorial plane fully analytically. 
We show that the unconditional upper limit
of the energy of the emitted particle
is given by 218.6 \% of that of the injected critical particle, 
irrespective of the details of the reaction
and this upper limit can be realized for massless particle emission.
The upper limit of the energy 
extraction efficiency for this emission 
as a collisional Penrose process 
is given by 146.6 \%, which 
can be realized in the collision of two massive particles with 
optimized mass ratio. 
Moreover, we analyze perfectly elastic collision,
Compton scattering, and pair annihilation and show that 
net positive energy extraction is really possible 
for these three reactions.
The Compton scattering is most efficient among them
and the efficiency can reach 137.2 \%.
On the other hand, our result is qualitatively consistent with the 
earlier claim that the mass and energy of the emitted particle 
are at most of order 
the total energy of the injected particles and hence
we can observe neither super-heavy nor super-energetic particles.
The present paper places the baseline for the study of 
particle emission from high-energy collision near a 
rapidly rotating black hole.

\end{abstract}

\maketitle


\newpage

\section{Introduction}
Ba\~nados, Silk, and West (2009)~\cite{Banados:2009pr} 
have indicated rapidly rotating Kerr black holes
as particle accelerators based on the demonstration that the center-of-mass (CM) energy of 
two colliding particles can be arbitrarily high near the horizon of a maximally rotating Kerr black hole if the angular momentum of either of the particles is finely tuned. Hereafter, we refer to this process as {\it Ba\~nados-Silk-West (BSW) process} or {\it BSW collision}.
In fact, the collision with infinite CM energy has already been 
noticed by Piran, Shaham, and Katz (1975)~\cite{Piran_etal_1975,Piran_Shaham_1977_upper_bounds,Piran_Shaham_1977_grb} in the study of 
an energy extraction process by two colliding particles in the 
ergo region, which is called a {\it collisional Penrose process}.
Recently, the particle acceleration by Kerr black holes has been investigated in different respects~\cite{Berti:2009bk,Jacobson:2009zg,Grib_Pavlov2010_Kerr,Grib:2010xj,Harada:2010yv,Harada:2011xz,Harada:2011pg,Zaslavskii:2011dz}, while 
it turns out that this phenomenon can be regarded as one of the 
general properties of extremal and 
near-extremal black 
holes~\cite{Zaslavskii2010_rotating,Zaslavskii2010_charged,Wei:2010vca,Liu:2011wv,Wei:2010gq,Yao:2011ai,Zhu:2011ae,Frolov:2011ea,Igata:2012js,
Kimura:2010qy} and other gravitating 
objects which are near-extremal in some specific
sense~\cite{Patil:2011ya,Patil:2011yb,Patil:2011aa,Patil:2012fu,Patil:2011uf}.

As for observability, we need to consider the emission from the BSW process.
The observed flux and characteristic spectrum from the pair annihilation of dark matter particles through 
the BSW collision around a Kerr black hole have been demonstrated in Refs.~\cite{Banados:2010kn,Williams:2011uz}.  
Since the collision with high CM energy can produce 
very massive particles, one might expect highly 
energetic particles can escape to infinity and 
be observed by a distant observer as the black hole is fed with 
product counterparts with largely negative energy.
On the other hand, Jacobson and 
Sotiriou (2010)~\cite{Jacobson:2009zg}
have claimed that for the collision of two particles 
of equal mass $m_{0}$, an ejecta particle cannot be 
more energetic than $2m_{0}$ and the energy upper limit of the ejecta tends to 
$m_{0}$ in the limit of infinite CM energy.
If this were the case, the BSW process would not be 
applicable to a collisional Penrose process.

In the present paper, we give the general formulation for the BSW collision 
and subsequent reaction. Based on this, we study the mass and energy of the particle
which escapes to infinity and 
obtain the unconditional upper limits of its mass and energy. 
We further derive the upper limit of the energy extraction efficiency
for this upper limit of energy emission
as a collisional Penrose process. We find that net positive energy extraction is really possible, although 
the efficiency is not very high but modest. 
We also study the upper limits of 
the energy of the emitted particle 
for specific physical processes and find that 
the energy extraction 
is really possible.
In summary, although the BSW process can be an applicable energy extraction mechanism 
to a collisional Penrose process, the mass and 
energy of the particles observable
to a distant observer are at most 
of order the total energy of the 
injected particles.

We use the units in which $G=c=1$ and follows the abstract index notation by Wald~\cite{Wald:1984rg}.

\section{Geodesic orbit, collision and reaction}

\subsection{Preliminaries} 
The line element in the Kerr spacetime 
in the Boyer-Lindquist coordinates is written in the following form~\cite{Kerr1963,Wald:1984rg,Poisson2004}:   
\begin{eqnarray*}
ds^{2}&=&-\left(1-\frac{2Mr}{\rho^{2}}\right)dt^{2}
-\frac{4Mar\sin^{2}\theta}{\rho^{2}}d\phi dt
+\frac{\rho^{2}}{\Delta}dr^{2}+\rho^{2}d\theta^{2} \nonumber \\
&&+\left(r^{2}+a^{2}+\frac{2Mra^{2}\sin^{2}\theta}{\rho^{2}}\right)
\sin^{2}\theta d\phi^{2} ,
\end{eqnarray*}
where $a$ and $M$ are the spin and mass parameters, respectively,
$\rho^{2}(r,\theta)=r^{2}+a^{2}\cos^{2}\theta$ and $\Delta(r)=r^{2}-2Mr+a^{2}$.
If $0< a^{2}\le M^{2}$, 
$\Delta$ vanishes at $r=r_{\pm}=M\pm\sqrt{M^{2}-a^{2}}$, where 
$r=r_{+}$ and $r=r_{-}$ correspond to an event and 
Cauchy horizons, respectively. Here, we denote $r_{+}=r_{H}$. 
Later, we will focus on the extremal case $a=M$.

In this paper we concentrate on geodesic particles 
in the equatorial 
plane $\theta=\pi/2$.
For a particle of mass $m$, energy $E$, and angular momentum $L$,  
the components of the four-momentum are given by
\begin{eqnarray}
p^{t}&=&\frac{1}{\Delta}\left[\left(r^{2}+a^{2}+\frac{2Ma^{2}}{r}\right)E-\frac{2Ma}{r}L\right], \\
p^{\phi}&=& \frac{1}{\Delta}\left[\left(1-\frac{2M}{r}\right)
L+\frac{2Ma}{r}E\right], \\
p^{\theta}&=& 0, 
\end{eqnarray}
and 
\begin{equation}
\frac{1}{2}(p^{r})^{2}+ V(r)=0,
\label{eq:p2square}
\end{equation}
where $V(r)$ is the effective potential given by 
\begin{equation}
V(r)=-\frac{Mm^{2}}{r}+\frac{L^{2}-a^{2}(E^{2}-m^{2})}{2r^{2}}
-\frac{M(L-aE)^{2}}{r^{3}}-\frac{E^{2}-m^{2}}{2}.
\label{eq:effective_potential}
\end{equation}
For a massless particle, we only have to choose $m=0$
in the above. For a massive particle, the four-velocity $u^{a}$, which is normalized as $u^{a}u_{a}=-1$,  
is given by $u^{a}=p^{a}/m$. 
The forward-in-time condition $p^{t}>0$ gives
\begin{equation}
\frac{1}{\Delta}\left[\left(r^{2}+a^{2}+\frac{2Ma^{2}}{r}\right)E-\frac{2Ma}{r}L\right]>0.
\label{eq:forward-in-time}
\end{equation}
In particular, this condition in the vicinity of the horizon $r\to r_{H}+0$ reduces to 
\begin{equation}
E-\Omega_{H}L\ge 0,
\end{equation}
where $\Omega_{H}=a/(r_{H}^{2}+a^{2})$ is the angular velocity of the horizon. We call $L_{c}=E/\Omega_{H}$ a critical angular momentum and a particle with this value of angular momentum a critical particle.

\subsection{Escape to infinity}
Next we discuss the escape of a particle to infinity
based on the effective potential.
First we consider massless particles. 
Solving $V(r)=0$ for the impact parameter $b=L/E$, we obtain
\begin{equation}
b=b_{\pm}(r)=\frac{-2aM \pm r \sqrt{\Delta(r) }}{r-2M}.
\end{equation}
This means that a massless particle with impact parameter $b=b_{\pm} (r)$ has a turning point at $r$. In particular, for $a=M$, we have
\begin{equation}
b_{+}(r)=r+M,\quad b_{-}(r)= -\left(r+M+\frac{4M^{2}}{r-2M}\right).
\end{equation}
$b_{+}(r)$ begins with $2M$ and monotonically 
increases to infinity as $r$ increases from $M$ to infinity. 
$b_{-}(r)$ begins with $2M$, is larger than $b_{+}(r)$, and monotonically 
increases to infinity as $r$ increases from $M$ to $2M$. 
As $r$ increases beyond $2M$ to infinity, 
$b_{-}(r) $ begins with negative infinity, monotonically increases 
to a local maximum $-7M $ at $r=4M$, and monotonically decreases to 
negative infinity. 
Thus, for $-7M<b<2M$, the particle escapes to infinity if it is 
moving outwardly initially. 
For $b>2M$ or $b<-7M$, the particle eventually escapes to infinity irrespective of the sign of the initial velocity if it is outside the 
outer turning point. 
For $b=2M$ or $b=-7M$, the particle escapes infinity, 
if it is moving outwardly outside the turning point, 
initially.
In other cases, the particle cannot escape to infinity. 

For massive particles, the situation is similar
except for energy dependence.
For convenience, we define $e=E/m$ and $\ell=L/(mM)$. 
For a massive particle, 
solving $V(r)=0$ for $\ell $, we obtain
\begin{equation}
\ell =\ell_{\pm}(r)=\frac{-2aMe \pm r \sqrt{\Delta(r)[(e^{2}-1)+2M/r]}}{M(r-2M)}.
\end{equation}
This means that a massive particle with angular momentum $\ell =\ell_{\pm} (r)$ has a turning point at $r$.
For bound particles, i.e. $e<1$, $V(r)$ becomes positive as $r$ goes sufficiently large, indicating that they cannot reach infinity but bounce back inwardly. Therefore, we concentrate on marginally bound and 
unbound particles, i.e. $e\ge 1$. 
For the maximal rotation $a=M$, 
$\ell_{+}(r)$ begins with $2e$ and monotonically increases to infinity as $r$ increases from $M$ to infinity. $\ell_{-}(r)$ begins with $2e$, is larger than $\ell_{+}(r)$, and monotonically increases to infinity as $r$ increases from $M$ to $2M$. As $r$ increases beyond $2M$ to infinity, $\ell_{-}(r)$ begins with negative infinity, monotonically increases to a negative local maximum value $\ell_{-,{\rm max}}(e)$, and then monotonically decreases to negative infinity. This means that the particle with $\ell$ satisfying $\ell_{L}(e)<\ell<\ell_{R}(e)$, where $\ell_{R}(e)=2e$ and $\ell_{L}(e)=\ell_{-,{\rm max}}(e)$,
escapes to infinity if it is moving outwardly initially. If the particle with $\ell$ satisfying $\ell >\ell_{R}(e)$ or $\ell<\ell_{L}(e)$ is outside the outer turning point, it eventually escapes to infinity irrespective of the sign of the initial radial velocity. 
For $\ell=\ell_{L}(r)$ or $\ell=\ell_{R}(r)$, the particle
escapes to infinity, if it is moving outwardly outside the 
turning point initially.
In other cases, the particle cannot escape to infinity. 

\subsection{Particle collision and reaction}
Here we consider the reaction of particles 1 and 2 into 3 and 4. 
We assume geodesic motion of each particle.
The local conservation of four-momentum before and after 
the collision is given by 
\begin{equation}
p_{1}^{\mu}+p_{2}^{\mu}=p_{3}^{\mu}+p_{4}^{\mu}.
\end{equation}
$\mu=t$ and $\mu=\phi$ yield the conservations of energy and 
angular momentum before and after the collision, i.e.
\begin{equation}
E_{1}+E_{2}= E_{3}+E_{4}, 
\label{eq:energy_conservation}
\end{equation}
and 
\begin{equation}
L_{1}+L_{2}= L_{3}+L_{4},
\label{eq:angular_momentum_conservation}
\end{equation}
respectively. $\mu=r$ yields 
\begin{equation}
p^{r}_{1}+p^{r}_{2}=p^{r}_{3}+p^{r}_{4}.
\label{eq:pr_conservation}
\end{equation}
Given incident particles 1 and 2, if we specify $m_{3}$, $E_{3}$ and $L_{3}$, 
we can determine
$m_{4}$, $E_{4}$, and $L_{4}$. In fact, 
$m_{4}$ can be expressed in terms of the 
quantities of other three particles as follows:
\begin{equation}
m_{4}^{2}=-p_{4a}p_{4}^{a}=-(p_{1}^{a}+p_{2}^{a}-p_{3}^{a})
(p_{1a}+p_{2a}-p_{3a}).
\end{equation}
On the other hand, the CM energy of particles 1 and 2 is given by
\begin{equation}
E_{\rm cm}^{2}=-(p_{1}^{a}+p_{2}^{a})(p_{1a}+p_{2a}).
\end{equation}
From the energy conservation, the total rest 
mass of product particles 3 and 4 must be smaller than or equal to the CM energy, i.e.
\begin{equation}
m_{3}+m_{4}\le E_{\rm cm}.
\label{eq:total_mass}
\end{equation}

The BSW process is characterized by 
$\tilde{L}_{1}= 2E_{1}$, $\tilde{L}_{2}<2E_{2}$, and 
$r\approx M$ for a maximally rotating black hole $a=M$,
where we have put $\tilde{L}=L/M$ for brevity.
The CM energy in this special case is derived 
in Refs.~\cite{Banados:2009pr,Jacobson:2009zg,Grib_Pavlov2010_Kerr,Grib:2010xj,Harada:2010yv}
in an explicit form as follows:
\begin{equation}
E_{\rm cm}\approx
\sqrt{\frac{2(2E_{1}-\sqrt{3E_{1}^{2}-m_{1}^{2}})(2E_{2}-\tilde{L}_{2})}{\epsilon}},
\end{equation}
where we denote the radius of the collision point 
as $r=M/(1-\epsilon)$ and $0<\epsilon\ll 1$.
For a critical particle, $E_{1}>m_{1}/\sqrt{3}$ must be satisfied.
As $\epsilon \to 0 $, the CM energy is diverging. 

\section{Collision and reaction near the horizon}
\subsection{Collision and reaction on the horizon}

From now on, we assume that the black hole is maximally rotating or $a=M$. We first consider the collision at $r=r_{H}=M$.
We assume that particle 1 is critical, while particle 2 is 
subcritical, i.e. $\tilde{L}_{1}=2E_{1}$ and $\tilde{L}_{2}<2E_{2}$.  
Note that although the collision we consider here is unphysical
because it takes infinite proper time for particle 1 to reach the horizon, it helps us to consider 
physical processes later. 
The forward-in-time condition on the horizon 
for particles 3 and 4 gives
\begin{equation}
2E_{3}-(2E_{2}-\tilde{L}_{2})\le \tilde{L}_{3}\le 2E_{3}.
\end{equation}

On the horizon $r=r_{H}=M$, from Eqs.~(\ref{eq:p2square}) and 
(\ref{eq:effective_potential}), we obtain
\begin{equation}
p^{r}=\sigma \left(2E-\tilde{L}\right),
\label{eq:pr_horizon}
\end{equation}
where $\sigma$ is the sign of $p^{r}$ and we have taken 
the forward-in-time condition into account to open 
the square root. 
Using Eq.~(\ref{eq:pr_horizon}), we can show that the left-hand 
side of Eq.~(\ref{eq:pr_conservation}) becomes
\begin{equation}
\sigma_{2}\left(2E_{2}-\tilde{L}_{2}\right),
\end{equation}
where we choose $\sigma_{2}=-1$. 
The right-hand side of Eq.~(\ref{eq:pr_conservation}) is
\begin{equation}
\sigma_{3}\left(2E_{2}-\tilde{L}_{2}\right)
\end{equation}
for $\sigma_{3}=\sigma_{4}$, while it is  
\begin{equation}
\sigma_{3}\left[4E_{3}-2E_{2}-(2\tilde{L}_{3}-\tilde{L}_{2})\right]
\end{equation}
for $\sigma_{3}=-\sigma_{4}$.
Hence, we can conclude $\sigma_{3}=\sigma_{4}=-1$ for 
the former case, while 
\begin{equation}
2E_{3}-\tilde{L}_{3}=0
\end{equation}
for $\sigma_{3}=1$,
and 
\begin{equation}
2E_{3}-\tilde{L}_{3}=2E_{2}-\tilde{L}_{2}
\end{equation}
for $\sigma_{3}=-1$ for the latter case. 

Note that particle 3 cannot leave the black hole
because it is released on the horizon. 
It is natural to introduce a reaction in the vicinity of the 
horizon as a small perturbation of the on-horizon reaction. 
It is clear that
we should concentrate on the case $\tilde{L}_{3}=2E_{3}$, otherwise 
there is no chance for a distant observer to observe particle 3 
even if the collision is slightly perturbed. This fixes 
$\sigma_{4}=-1$. 

\subsection{Near-horizon behavior of particles}
We consider a collision near the horizon, where 
$r=M/(1-\epsilon)$ and $\tilde{L}_{3}=2E_{3}(1+\delta)$, where
$0< \epsilon\ll 1$ and $|\delta|\ll 1$.
We assume that $\delta$ can be expanded in powers 
of $\epsilon$ as follows:
\begin{equation}
\delta=\delta_{(1)}\epsilon+\delta_{(2)}\epsilon^{2}+O(\epsilon^{3}).
\end{equation}
This assumption will be justified later because it gives a 
consistent expansion of the four-momentum conservation.

\begin{center}
\begin{figure}[htbp]
\begin{tabular}{lr}
\includegraphics[width=0.45\textwidth]{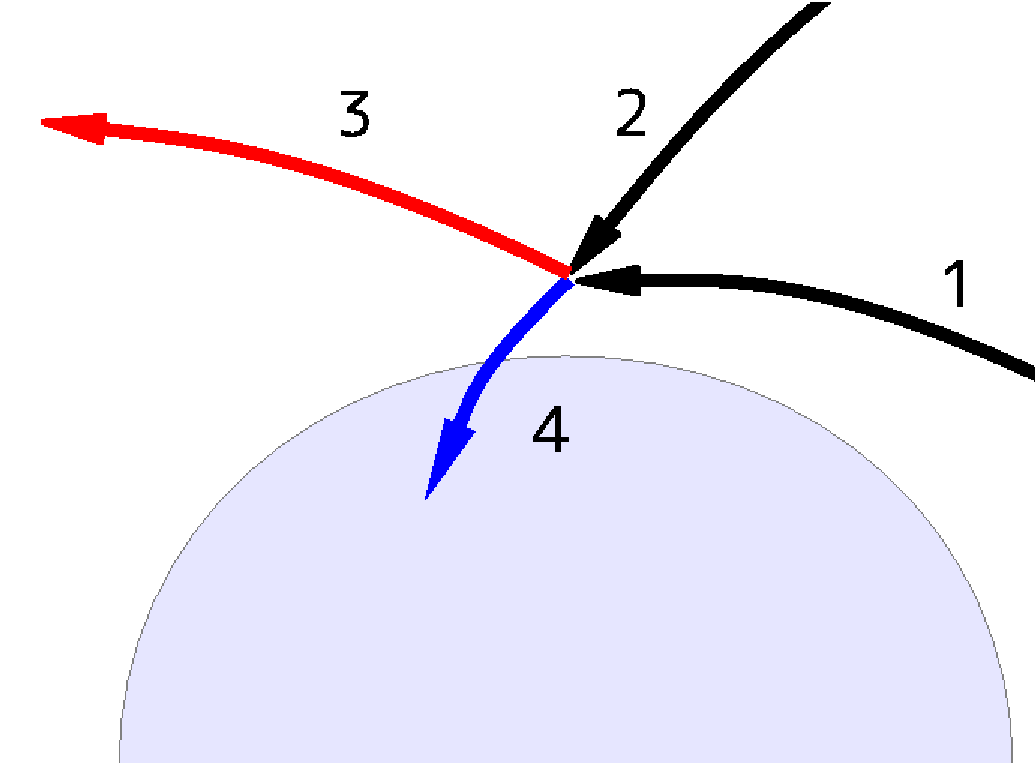}&
\includegraphics[width=0.45\textwidth]{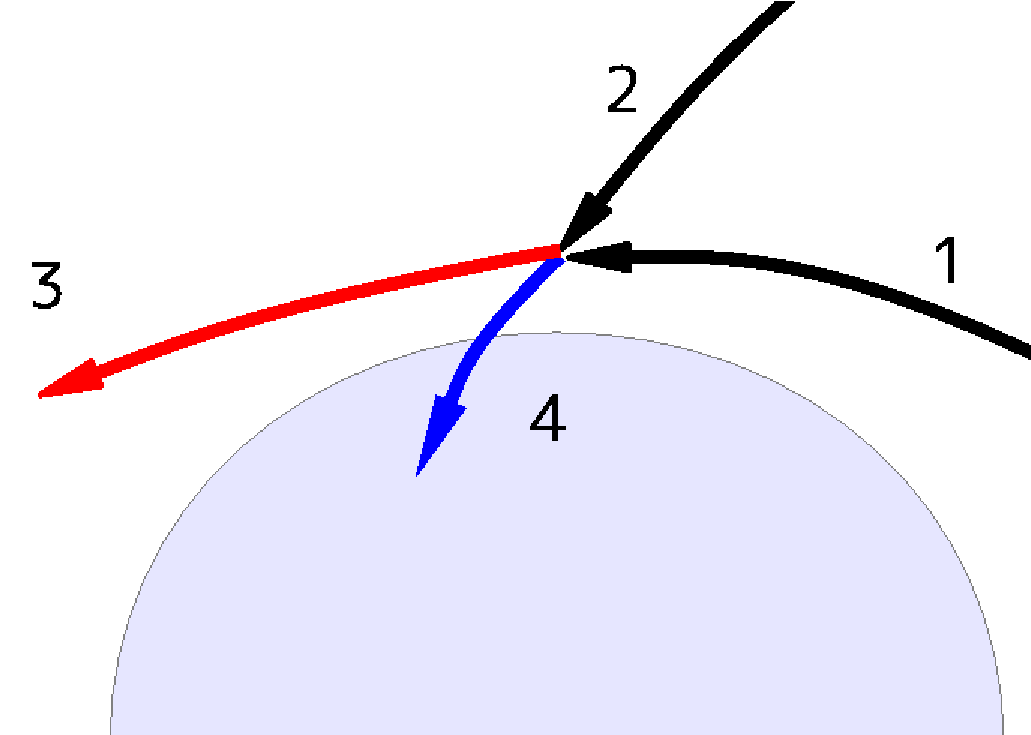}
\end{tabular}
\caption{\label{fg:reaction} The left and right panels are the schematic figures of reactions, where particle 3 has outward ($\sigma_{3}=1$) and inward ($\sigma_{3}=-1$) initial velocities, respectively.
}
\end{figure}
\end{center}

Here, we require particle 3 to escape to infinity.
This is possible in the following two cases: (a) $e_{3}\ge 1$, 
$\ell_{L}(e_{3})<\ell_{3}\le \ell_{R}(e_{3})$, 
and $\sigma_{3}=1$
and (b) $e_{3}\ge 1$, $\ell_{3}>\ell_{R}(e_{3})$, and $r\ge r_{t,+}(e_{3})$, where 
$r_{t,+}(e)$ is the radius of the outer turning point for a particle with $e$.
The left and right panels of Fig.~\ref{fg:reaction} 
give the schematic figures for $\sigma_{3}=1$ and $-1$, respectively.
The reason why the case $\ell_{3}\le \ell_{L}(e_{3})$ 
is not considered is that the two turning points are both well outside the horizon in this case.

Under these conditions, we will see the upper limits of the
mass $m_{3}$ and energy $E_{3}$ of particle 3. 
Since $\ell_{R}(e)=2e$ for a maximally rotating Kerr black hole, 
we need to have $\delta\le 0$ and $\sigma_{3}=1$ for case (a). For case (b), since the turning points are given by
\begin{equation}
r_{t,\pm}(e)=M\left(1+\frac{2e}{2e\mp\sqrt{e^{2}+1}}\delta_{(1)}\epsilon\right)+O(\epsilon^{2}),
\label{eq:turning_points}
\end{equation} 
$r\ge r_{t,+}(e)$ implies
\begin{equation}
0\le \delta_{(1)}\le \frac{2E_{3}-\sqrt{E_{3}^{2}+m_{3}^{2}}}{2E_{3}}=\delta_{(1),{\rm max}}.
\label{eq:xrange_caseb}
\end{equation}
Note that the forward-in-time condition Eq.~(\ref{eq:forward-in-time}) onto particle 3 in the vicinity of the horizon reduces to 
\begin{equation}
\delta<\epsilon+\frac{7}{4}\epsilon^2+O(\epsilon^3).
\end{equation}
Therefore, $\delta_{(1)}<1$ gives a sufficient condition and 
this is already guaranteed for both cases (a) and (b).

We can easily show that the above argument applies 
for massless particles by the appropriate 
replacement of $\ell$ with $b$
and taking the limit $m\to 0$ and $e\to \infty$ 
in Eqs.~(\ref{eq:turning_points}) and (\ref{eq:xrange_caseb}).
\subsection{Local momentum conservation}
To look into the local momentum conservation, we use a series of 
$|p^{r}|$ in powers of $\epsilon$ for each particle as follows: 
\begin{eqnarray}
|p_{1}^{r}|&=&\sqrt {3E_{1}^{2}-m_{1}^{2}}\epsilon-\frac{E_{1}^{2}}{\sqrt{3E_{1}^{2}-m_{1}^{2}}}\,{\epsilon}^{2}
+O(\epsilon^{3}), 
\label{eq:p1r_series}\\
|p_{2}^{r}|&=& \left(2E_{2}-\tilde{L}_{2} \right) +2\, \left(\tilde{L}_{2}-E_{2} \right) \epsilon +\frac {{\tilde{L}_{2}}^{2}-4\,\tilde{L}_{2}E_{2}+3E_{2}^{2}-m_{2}^{2}}{2(2E_{2}-\tilde{L}_{2})}\epsilon^{2}+O(\epsilon^{3}), \\
|p_{3}^{r}|&=&\sqrt {{E_{3}}^{2}(3-8\delta_{(1)}+4\delta_{(1)}^{2})-m_{3}^{2}}\epsilon -{\frac {{E_{3}}^{
2} \left[ 1-4(2\delta_{(1)}-\delta_{(2)})(1-\delta_{(1)}) \right]}{\sqrt {E_{3}^{2}(3-8\delta_{(1)}+4\delta_{(1)}^{2})-m_{3}^{2}}}} {\epsilon}^{2} \nonumber \\
&& +O(\epsilon^{3}), \\
|p_{4}^{r}|&=& \left( 2E_{2}-\tilde{L}_{2} \right)+ 
\left[ 2 (\tilde{L}_{2}-E_{2})+2E_{3}\,(\delta_{(1)}-1)
+2E_{1}\right] \epsilon \nonumber \\
&&+\left[\frac{(2E_{2}-\tilde{L}_{2})}{2}-2(2\delta_{(1)}-\delta_{(2)})E_{3}
-\frac{(E_{1}+E_{2}-E_{3})^{2}+m_{4}^{2}}{2(2E_{2}-\tilde{L}_{2})}
\right]\epsilon^{2}
+O(\epsilon^{3}) ,
\end{eqnarray}
where in the last equation we have used 
Eqs.~(\ref{eq:energy_conservation}) and (\ref{eq:angular_momentum_conservation}) 
to eliminate $E_{4}$ and $\tilde{L}_{4}$. 

The first and second order terms of $\epsilon$ in Eq.~(\ref{eq:pr_conservation}) then give
\begin{equation}
(2E_{1}-\sqrt{3E_{1}^{2}-m_{1}^{2}})+2E_{3}(\delta_{(1)}-1)
=\sigma_{3}\sqrt{E_{3}^{2}(3-8\delta_{(1)}+4\delta_{(1)}^{2})-m_{3}^{2}}
\label{eq:general_1st_order}
\end{equation}
and
\begin{eqnarray}
&& \frac{E_{1}^{2}}{\sqrt{3E_{1}^{2}-m_{1}^{2}}}+\frac{\tilde{L}_{2}^{2}-4 \tilde{L}_{2}E_{2}+3E_{2}^{2}-m_{2}^{2}}{2(\tilde{L}_{2}-2E_{2})}=
-\sigma_{3}\frac{E_{3}^{2}[1-4(2\delta_{(1)}-\delta_{(2)})(1-\delta_{(1)})]}{\sqrt{E_{3}^{2}(3-8\delta_{(1)}+4\delta_{(1)}^{2})-m_{3}^{2}}} \nonumber \\
&& \quad -\left[\frac{(2E_{2}-\tilde{L}_{2})}{2}-2(2\delta_{(1)}-\delta_{(2)})E_{3}
-\frac{(E_{1}+E_{2}-E_{3})^{2}+m_{4}^{2}}{2(2E_{2}-\tilde{L}_{2})}
\right],
\label{eq:general_2nd_order}
\end{eqnarray}
respectively.

\section{Unconditional upper limits for general reaction \label{sec:unconditional}}
\subsection{Mass and energy of the emitted particle\label{subsec:mass_energy}}

Taking the square of the both sides of 
Eq.~(\ref{eq:general_1st_order}), we can derive
\begin{equation}
1-\delta_{(1)}=\frac{A_{1}^{2}+(E_{3}^{2}+m_{3}^{2})}{4 A_{1}E_{3}},
\label{eq:general_x}
\end{equation}
where we put $A_{1}=2E_{1}-\sqrt{3E_{1}^{2}-m_{1}^{2}}>0$ for convenience. Note that Eq.~(\ref{eq:general_x}) immediately implies
\begin{equation}
\delta_{(1),{\rm max}}-\delta_{(1)}=\frac{(A_{1}-\sqrt{E_{3}^{2}+m_{3}^{2}})^{2}}{4A_{1}E_{3}}\ge 0.
\end{equation}

First we consider case (a).
Substituting Eq.~(\ref{eq:general_x}) into the left-hand side of Eq.~(\ref{eq:general_1st_order}), 
we obtain
\begin{equation}
A_{1}-\frac{E_{3}^{2}+m_{3}^{2}}{A_{1}}=2\sigma_{3}\sqrt{E_{3}^{2}(3-8\delta_{(1)}+4\delta_{(1)}^{2})-m_{3}^{2}}.
\label{eq:general_1st_order_variant}
\end{equation}
Since $\sigma_{3}=1$, Eq.~(\ref{eq:general_1st_order_variant}) implies
\begin{equation}
A_{1}^{2}-(E_{3}^{2}+m_{3}^{2})
\ge 0.
\label{eq:general_sigma3=1}
\end{equation}
This implies $m_{3}\le A_{1}$ and  
\begin{equation} 
E_{3}\le \sqrt{A_{1}^{2}-m_{3}^{2}}=\lambda_{0}.
\end{equation}
Since $\lambda_{0}\le (2-\sqrt{2})E_{1}$ for $E_{1}\ge m_{1}$, which we assume as the injection 
of particle 1 from infinity to the system,  
we cannot extract net positive energy with $\sigma_{3}=1$.

For case (b), only the range given by Eq.~(\ref{eq:xrange_caseb}) 
is permitted. Although both $\sigma_{3}=\pm 1$ are possible, 
we cannot extract net positive energy for $\sigma_{3}=1$ 
as we have already shown.
So, we will concentrate on 
the case $\sigma_{3}=-1$. 
Equation~(\ref{eq:general_1st_order_variant}) then implies 
\begin{equation}
E_{3}^{2}\ge A_{1}^{2}-m_{3}^{2}.
\label{eq:sigma3_negative}
\end{equation}
$\delta_{(1)}\ge 0$ in 
Eq.~(\ref{eq:general_x}) implies
\begin{equation}
A_{1}^{2}+(E_{3}^{2}+m_{3}^{2})-4A_{1}E_{3}\le 0.
\label{eq:general_E3_inequality_negativex}
\end{equation}
The discriminant $D$ and roots $\lambda_{\pm}$ of the left-hand side
of Eq.~(\ref{eq:general_E3_inequality_negativex}) as a quadratic of $E_{3}$ are given by
\begin{equation}
D/4=3A_{1}^{2}-m_{3}^{2}
\end{equation}
and 
\begin{equation}
\lambda_{\pm}=2A_{1}\pm \sqrt{3A_{1}^{2}-m_{3}^{2}},
\label{eq:lambda_pm}
\end{equation}
respectively.
The solution of Eq.~(\ref{eq:general_E3_inequality_negativex}) 
is given by
\begin{equation}
\lambda_{-}\le E_{3}\le \lambda_{+},
\label{eq:unconditional_upper_bound}
\end{equation}
where $D\ge 0$ or $m_{3}\le \sqrt{3}A_{1}$ must be satisfied.
To have $E_{3}\ge m_{3}$, we need $\lambda_{+}\ge m_{3}$, for which
$m_{3}\le A_{1}/(2-\sqrt{2})$ must be satisfied.
The condition~(\ref{eq:sigma3_negative}) is satisfied
for $E_{3}=\lambda_{+}$ trivially if $m_{3}\ge A_{1}$ 
and because $\lambda_{+}>\lambda_{0}$ if $0\le m_{3}< A_{1}$. 
$\delta_{(1)}=0$ holds for $E_{3}=\lambda_{\pm}$.
We should note that $\lambda_{+}=E_{1}$ if both particles 1 and 
3 are massless, $\lambda_{+}<E_{1}$ if particles 1 and 3 are massless and massive, respectively, but $\lambda_{+}>E_{1}$ if particles 1 and 3 
are massive and massless, respectively.
Since $\lambda_{+}\le E_{1}$ in the limit $E_{1}/m_{1}\to \infty $, no net positive energy extraction is possible if the incident critical particle is highly energetic or massless.

\begin{center}
\begin{figure}[htbp]
\begin{tabular}{lr}
\includegraphics[width=0.45\textwidth]{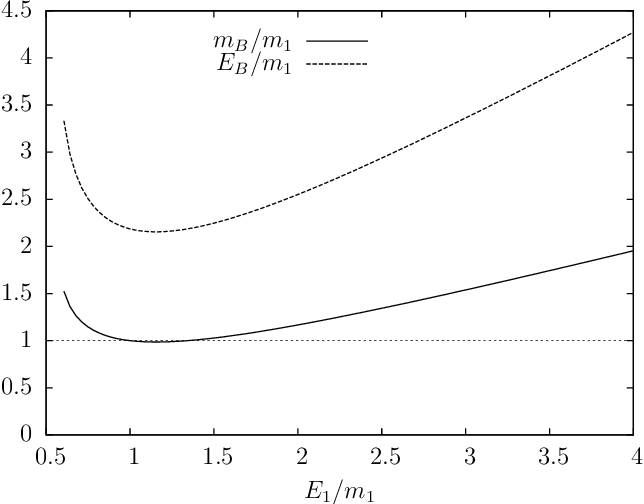}&
\includegraphics[width=0.45\textwidth]{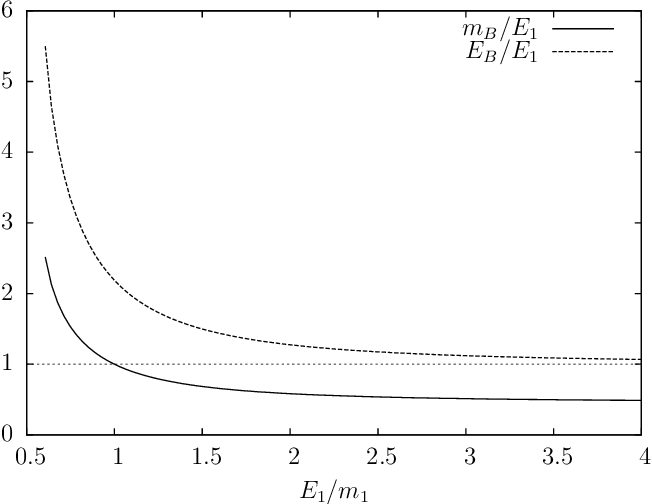}
\end{tabular}
\caption{\label{fg:upper_bounds} Upper limits of the mass and energy of the emitted particle as functions of the energy of the incident critical particle. The left and right panels show 
the ratios to the mass and to the energy of the incident critical particle, respectively. The mass and energy upper limits are denoted by the solid and dashed lines, respectively.}
\end{figure}
\end{center}

From the above argument, 
the unconditional upper limits of the mass and energy 
of the emitted particle 3 are given by
\begin{equation}
m_{3}\le (2E_{1}-\sqrt{3E_{1}^{2}-m_{1}^{2}})/(2-\sqrt{2})=m_{B}
\end{equation}
and 
\begin{equation}
E_{3}\le (2E_{1}-\sqrt{3E_{1}^{2}-m_{1}^{2}})/(2-\sqrt{3})=E_{B},
\end{equation}
respectively. 
Note that $\lambda_{+}=E_{B}$ can be realized only if particle 3 is massless. 
Figure~\ref{fg:upper_bounds} shows the upper limits as functions 
of $E_{1}/m_{1}$. 
$m_{B}/m_{1}=1$ at $E_{1}/m_{1}=1$ and $7-4\sqrt{2}$ and $m_{B}/m_{1}$
takes a minimum $(2+\sqrt{2})/(2\sqrt{3})\simeq 0.9856$ at $E_{1}/m_{1}=2/\sqrt{3}$.
$E_{B}/m_{1}=(2+\sqrt{3})(2-\sqrt{2})\simeq 2.186$ 
at $E_{1}/m_{1}=1$ and $7-4\sqrt{2}$  and $E_{B}/m_{1}$ takes a minimum $1+2/\sqrt{3}\simeq 2.154$ at $E_{1}/m_{1}=2/\sqrt{3}$. 
On the other hand, both $m_{B}/E_{1}$ and $E_{B}/E_{1}$ 
monotonically decrease as $E_{1}/m_{1}$ increases. $m_{B}/E_{1}$ 
takes a maximum 1 at $E_{1}/m_{1}=1$ and approaches $(2-\sqrt{3})/(2-\sqrt{2})\simeq 
0.4574$ as $E_{1}/m_{1}$ increases from 1 to infinity.
$E_{B}/E_{1}$ takes a maximum $(2+\sqrt{3})(2-\sqrt{2})\simeq 2.186$ 
at $E_{1}/m_{1}=1$ 
and approaches 1 as $E_{1}/m_{1}$ increases from 1 to infinity.
The mass and energy of the 
emitted particle can be at most of order the energy 
of the incident critical particle.
The upper limit $m_{B}$ of the mass of the emitted particle
is approximately equal to $m_{1}$ for $E_{1}\simeq m_{1}$
but can be much larger than $m_{1}$ for $E_{1}\gg m_{1}$. 
Since $E_{B}>E_{1}$,  
we might obtain the energy of the ejecta particle more than 
the total energy of the injected particles.
This possibility will be investigated in Sec.~\ref{subsec:energy_efficiency}.

\subsection{Energy extraction efficiency\label{subsec:energy_efficiency}}

Equation~(\ref{eq:general_2nd_order}) can be solved for $m_{4}^{2}$ as follows:
\begin{eqnarray}
m_{4}^{2}&=&(2E_{2}-\tilde{L}_{2})\left[\frac{2E_{1}^{2}}{\sqrt{3E_{1}^{2}-m_{1}^{2}}}
-4(2\delta_{(1)}-\delta_{(2)})E_{3}
+2\sigma_{3}\frac{E_{3}^{2}[1-4(2\delta_{(1)}-\delta_{(2)})(1-\delta_{(1)})]}
{\sqrt{E_{3}^{2}(3-8\delta_{(1)}+4\delta_{(1)}^{2})-m_{3}^{2}}}\right] \nonumber \\
&&+(E_{2}^{2}+m_{2}^{2})-(E_{1}+E_{2}-E_{3})^{2}.
\label{eq:m4_general}
\end{eqnarray}
Since $\delta_{(1)}$ is given by Eq.~(\ref{eq:general_x}), we can obtain $\delta_{(2)}$ using $m_{3}$ and $E_{3}$ for given $m_{4}$.
$E_{4}$ and $\tilde{L}_{4}$ are given by 
Eqs.~(\ref{eq:energy_conservation}) and (\ref{eq:angular_momentum_conservation}). For the collision to occur, $m_{4}^{2}\ge 0$ must 
be satisfied.
We should note that since $m_{2}$, $E_{2}$, $\tilde{L}_{2}$ and $\delta_{(2)}$, which do not appear in Eq.~(\ref{eq:general_x}), do appear in Eq.~(\ref{eq:m4_general}), the condition $m_{4}^{2}\ge 0$ can be generally satisfied.
Equation~(\ref{eq:m4_general}) seems to suggest that we can expect 
very large $m_{4}$ as $E_{1}\to m_{1}/\sqrt{3}$, although 
particle 4 cannot escape to infinity.
However, $E_{1}\to m_{1}/\sqrt{3}$ is a singular limit 
in the series of $|p_{1}^{r}|$ given by Eq.~(\ref{eq:p1r_series}). 
In Appendix~\ref{sec:circular_orbit}, we demonstrate that 
the apparently divergent term in this limit is replaced with 
a finite term for a particle circularly orbiting near the horizon.

In Sec.~\ref{subsec:mass_energy}, we have seen that 
the upper limit $E_{3}=\lambda_{+}$ can be realized 
only for $\delta_{(1)}= 0$. Here we show that 
this emission can be realized
and place the upper limit of the efficiency of the 
energy extraction for this emission.
The expression for $m_{4}$ is 
reduced to a simpler form for $\delta_{(1)}=0$ and 
$E_{3}=\lambda_{+}$ as follows:
\begin{eqnarray}
m_{4}^{2}&=&(2E_{2}-\tilde{L}_{2})\left[\frac{2E_{1}^{2}}{\sqrt{3E_{1}^{2}-m_{1}^{2}}}-\frac{2\lambda_{+}^{2}}
{\sqrt{3\lambda_{+}^{2}-m_{3}^{2}}}
-4\frac{2\lambda_{+}-\sqrt{3\lambda_{+}^{2}-m_{3}^{2}}}{\sqrt{3\lambda_{+}^{2}-m_{3}^{2}}}\lambda_{+}\delta_{(2)}
\right] \nonumber \\
&&+(E_{2}^{2}+m_{2}^{2})-(E_{1}+E_{2}-\lambda_{+})^{2}.
\label{eq:m4_general_lambda+}
\end{eqnarray}
This means that even if $\delta_{(1)}=0$, we can still have 
different values for $m_{4}$ by adjusting $\delta_{(2)}$.

As we have already seen, we can obtain net positive 
energy gain only for 
$\delta>0$. Since the upper limit $E_{3}=\lambda_{+}$ is obtained 
for $\delta_{(1)}= 0$, we need to assume $\delta_{(2)}\ge 0$. 
Then, since $r_{t,+}=M+O(\epsilon^2)$, 
the collision point $r=M/(1-\epsilon)$ is outside the outer turning point. 
Since $\lambda_{+}=E_{B}$ only if particle 3 is 
massless, we concentrate on this case.
In this case, we can prove that the first term 
on the right-hand side of Eq.~(\ref{eq:m4_general_lambda+}) 
is negative.
The condition for $E_{2}$ is then given by 
\begin{equation}
E_{2}\ge \frac{1}{2}\left[(\lambda_{+}-E_{1})-\frac{m_{2}^{2}}{\lambda_{+}-E_{1}}\right]=\kappa.
\label{eq:e2_lower_bound}
\end{equation}
The proof for this condition will be postponed 
until Appendix~\ref{sec:negative}.
This implies that $E_{2}$ cannot vanish but greater than or 
equal to $(\lambda_{+}-E_{1})/2$ even if particle 2 is massless and that particle 2 must be unbound if $\kappa>m_{2}$.
Conversely, we can always find $m_{4}$ and $\tilde{L}_{2}$ satisfying $m_{4}^{2}\ge 0$ and $\tilde{L}_{2}<2 E_{2}$ if the above inequality is satisfied.

Since Eq.~(\ref{eq:e2_lower_bound}) potentially gives a lower limit of $E_{2}$, this can constrain the efficiency of the energy extraction $\eta=E_{3}/(E_{1}+E_{2})$ for $E_{3}=\lambda_{+}$. To estimate $\eta$, we here assume $E_{2}\ge m_{2}$ as usual, i.e. we inject the two incident particles from infinity. If $\kappa >  m_{2} $ or $m_{2}< (\lambda_{+}-E_{1})/(\sqrt{2}+1)$, we find
\begin{equation}
\eta\le \frac{\lambda_{+}}{E_{1}+\kappa}=1+\frac{(\lambda_{+}-E_{1})^{2}+m_{2}^{2}}{\lambda_{+}^{2}-E_{1}^{2}-m_{2}^{2}}.
\label{eq:eta_max_1}
\end{equation}
Therefore, the upper limit exceeds unity and hence we can obtain net positive energy extraction.
If $\kappa \le  m_{2} $ or $m_{2}\ge (\lambda_{+}-E_{1})/(\sqrt{2}+1)$, we find
\begin{equation}
\eta\le \frac{\lambda_{+}}{E_{1}+m_{2}}.
\end{equation}
Hence, net positive energy extraction is possible 
with the upper limit if and only if $m_{2}<\lambda_{+}-E_{1}$.

\begin{center}
\begin{figure}[htbp]
\includegraphics[width=0.7\textwidth]{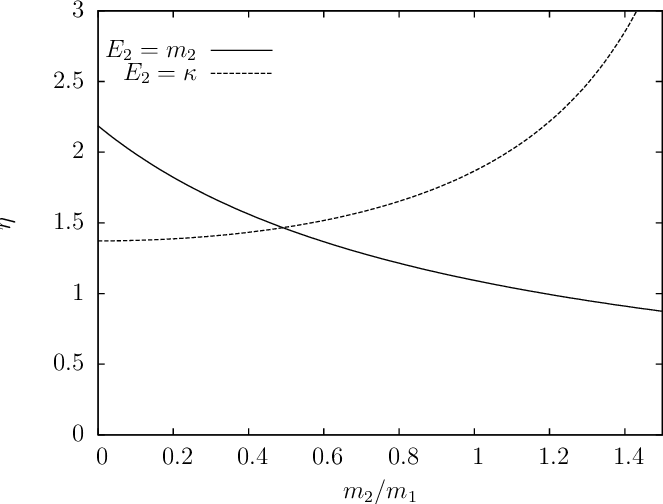}
\caption{\label{fg:efficiency} The upper limit of the 
energy extraction efficiency 
for the upper limit of ejecta energy $E_{3}=E_{B}$
as a function of the mass ratio $m_{2}/m_{1}$, where $E_{1}=m_{1}$ and $m_{3}=0$ are chosen. The solid and dashed lines denote the
efficiencies for $E_{2}=m_{2}$ and $E_{2}=\kappa$, respectively.
If the mass ratio is smaller than 0.4913, we should adopt $E_{2}=\kappa$, while if the ratio is greater than this value, we should adopt $E_{2}=m_{2}$. The efficiency takes a maximum 1.466 at $m_{2}/m_{1}=0.4913$, where the two curves cross each other.}
\end{figure}
\end{center}

We can here determine the unconditional upper limit of $\eta$
for $E_{3}=\lambda_{+}$.
Since $\lambda_{+}$ does not depend on $E_{2}$, 
to maximize the upper limit of $\eta$, we 
should first find the value for $m_{2}$
which minimizes $E_{2}$ for fixed $E_{1}$ and $m_{1}$.
This corresponds to 
the case $\kappa= m_{2}$ or $m_{2}= (\lambda_{+}-E_{1})/(\sqrt{2}+1)$.
$\eta$ is then maximized for $E_{1}=m_{1}$. Therefore, the unconditional upper limit is given by  
\begin{eqnarray*}
\eta_{B}=\frac{\lambda_{+}}{m_{1}+m_{2}}
=\frac{(\sqrt{2}+1)\lambda_{+}}{\sqrt{2}m_{1}+\lambda_{+}}.
\end{eqnarray*}
Since $\lambda_{+}$ takes the upper limit $E_{B}=(2+\sqrt{3})(2-\sqrt{2})m_{1}$ for $m_{3}=0$, the unconditional upper limit is given by  
\begin{eqnarray*}
\eta_{B}=  \frac{2+\sqrt{2}+\sqrt{6}}{4} \simeq 1.466,
\end{eqnarray*}
where 
$m_{2}/m_{1}=(5\sqrt{2}-4\sqrt{3}+3\sqrt{6}-7)\simeq 0.4913 $.
Because we impose the condition $E_{2}\ge m_{2}$, 
the upper limit of the efficiency is realized 
at the crossing point of the two curves, $E_{2}=\kappa$  and $E_{2}=m_{2}$. Figure~\ref{fg:efficiency}
shows the upper limit of $\eta$ as a function of the mass ratio $m_{2}/m_{1}$, where we choose particle 1 as marginally bound, i.e. $E_{1}=m_{1}$ for reference. It begins with $2(54-10\sqrt{2}+14\sqrt{3}+\sqrt{6})/97\simeq 1.372$
and monotonically increases to $(2+\sqrt{2}+\sqrt{6})/4\simeq 1.466$ as
$m_{2}/m_{1}$ increases from $0$ to $5\sqrt{2}-4\sqrt{3}+3\sqrt{6}-7\simeq 0.4912$, where $E_{2}=\kappa$. 
Then, the upper limit monotonically decreases to $0$ as $m_{1}/m_{2}$ increases beyond this value, where $E_{2}=m_{2}$. 
It becomes 
$(2+\sqrt{3})(2-\sqrt{2})/2\simeq 1.093$ at $m_{1}/m_{2}=1$. 

In the end of the general analysis, it should be noted that 
the present mass, energy, and efficiency upper limits of the emission from the BSW collision 
are applicable even if product particles are more than two. 
This is because in such cases we can regard more than one product particles 
other than particle 3 as those produced as a result of the decay of 
particle 4. Thus, the present upper limits are unconditional in the sense that they are applicable irrespective of the details 
of the incident counterpart and the product particles.

\section{Upper limits for specific physical reactions}
In this section, we specify physical 
reaction models and discuss the upper limits 
of the energy of the emitted particle and the energy extraction efficiency, based on the result obtained in Sec.~\ref{sec:unconditional}.

\subsection{Perfectly elastic collision}
We first consider perfectly elastic collision of equal masses, i.e.
$m_{1}=m_{2}=m_{3}=m_{4}=m_{0}$. 
We choose particle 1 as marginally bound for reference, i.e.
$E_{1}=m_{0}$.
Then, from Eq.~(\ref{eq:lambda_pm}), the upper limit of the energy of 
particle 3 is given by 
\begin{equation}
\lambda_{+}=(7-4\sqrt{2})m_{0}\simeq 1.343 m_{0},
\end{equation}
where $E_{3}=\lambda_{+}$ is realized for $\delta_{(1)}= 0$.
In fact, if $m_{1}=m_{3}$, we can easily prove that the 
first term on the right-hand side of Eq.~(\ref{eq:m4_general_lambda+})
is nonpositive because $\lambda_{+}\ge E_{1}$. Then, the argument
similar to that given in Sec.~\ref{subsec:energy_efficiency} 
applies. 
Since $m_{2}=m_{0}>(\lambda_{+}-E_{1})/(\sqrt{2}+1)$, we choose particle 2 as marginally bound, i.e. $E_{2}=m_{0}$, and hence
the upper limit of $\eta$ for $E_{3}=\lambda_{+}$ is given by
\begin{equation}
\eta\le \frac{\lambda_{+}}{2m_{0}}=\frac{7-4\sqrt{2}}{2}\simeq 0.6716.
\end{equation}
Therefore, we can obtain no net positive energy extraction.
The above result will be discussed later in direct comparison with the claim in Ref.~\cite{Jacobson:2009zg}. 

Next we assume that 
$m_{1}=m_{3}$ and $m_{2}=m_{4}$ but not $m_{1}=m_{2}$. 
For $E_{1}=m_{1}$, the upper limit of $E_{3}$ is given by 
\begin{equation}
\lambda_{+}=(7-4\sqrt{2})m_{1}.
\end{equation}
We can optimize $m_{2}$ to 
$m_{2}=(\lambda_{+}-E_{1})/(\sqrt{2}+1)=2(5\sqrt{2}-7)m_{1}\simeq 0.1421 m_{1}$ so that we can obtain the upper limit of the energy extraction efficiency for $E_{3}=\lambda_{+}$ as follows:
\begin{equation}
\frac{\lambda_{+}}{m_{1}+m_{2}}=\frac{18\sqrt{2}+11}{31}\simeq 1.176.
\end{equation}
Therefore, net positive energy extraction is possible for perfectly elastic collision if the mass of the counterpart is in some range. 
The upper limit of the energy extraction efficiency becomes 117.6 \%, 
where the mass ratio is optimized.

\subsection{Compton scattering}

We here assume that particle 3 is massless. This is motivated by the fact that the unconditional energy upper limit $E_{B}$ can be realized only if particle 3 is massless.
If we consider the Compton scattering, we can identify 
either of particles 1 and 2 with a massless particle.

First we assume particle 1 is massless and hence $m_{1}=m_{3}=0$
and $m_{2}=m_{4}=m_{0}$. Then, the upper limit of 
the energy of particle 3 is given by
\begin{equation}
\lambda_{+}=E_{1}.
\end{equation}
With $E_{3}=E_{1}$ and $\delta_{(1)}=0$, 
Eq.~(\ref{eq:m4_general_lambda+}) yields $\delta_{(2)}=0$.
In other words, up to this order particles 1 and 2 just passed through
each other and no energy nor angular momentum is exchanged. 
We cannot determine whether particle 3 can escape to infinity up to this order.
Even if particle 3 can escape to infinity, we have no net positive energy extraction anyway.

Next, we assume particle 2 is massless and hence $m_{1}=m_{4}=m_{0}$,
$m_{2}=m_{3}=0$, and $E_{1}=m_{0}$. In this case,
\begin{equation}
\lambda_{+}=(2+\sqrt{3})(2-\sqrt{2})m_{0}\simeq 2.186 m_{0}.
\end{equation}
Since $m_{2}=0<(\lambda_{+}-E_{1})/(\sqrt{2}+1)$, the upper limit of $\eta$ for $E_{3}=\lambda_{+}$ becomes
\begin{equation}
\eta\le 1+\frac{\lambda_{+}-E_{1}}{\lambda_{+}+E_{1}}=\frac{2(54-10\sqrt{2}+14\sqrt{3}+\sqrt{6})}{97}\simeq 1.372.
\end{equation}
This is comparable with the unconditional upper limit 1.466. 
The (inverse) Compton scattering between a subcritical photon and a critical 
massive particle is rather efficient as a collisional Penrose process.
\subsection{Pair annihilation}

We here consider pair annihilation of two equal masses into 
two massless particles. Then, $m_{1}=m_{2}=m_{0}$ and $m_{3}=m_{4}=0$. We additionally assume $E_{1}=m_{0}$ for reference. In this case, the upper limit of the energy of particle 3 is given by
\begin{equation}
\lambda_{+}=(2+\sqrt{3})(2-\sqrt{2})m_{0}\simeq 2.186 m_{0}.
\label{eq:pair_annihilation_energy_upper_limit}
\end{equation}
In this case, since $m_{2}=m_{0}>(\lambda_{+}-E_{1})/(\sqrt{2}+1)m_{0}$, we choose particle 2 as marginally bound, i.e. $E_{2}=m_{0}$, and hence 
the upper limit of $\eta$ is given by 
\begin{equation}
\eta\le \frac{(2+\sqrt{3})(2-\sqrt{2})}{2}\simeq  1.093.
\end{equation}
Thus, net 9.3 \% of the total injected energy can be extracted.
This result will be also discussed later in comparison 
with Ref.~\cite{Jacobson:2009zg}.

\section{Discussion and conclusion}

We have studied particle emission from the BSW collision and 
subsequent reaction,
where a critical particle 
collides with a generic counterpart particle 
near the horizon of a maximally rotating Kerr black hole.
Since the CM energy of the two particles 
can be arbitrarily high, the collision can produce 
very massive and/or energetic particles and one might 
speculate that such particles can potentially escape to infinity
through a collisional Penrose process.  
We have however found that this is not the case. We cannot 
observe a particle much more massive nor much more energetic 
than the energy of the incident critical particle. This is qualitatively consistent 
with the earlier results~\cite{Piran_etal_1975,Piran_Shaham_1977_upper_bounds,Piran_Shaham_1977_grb,Jacobson:2009zg}.

We have derived
the unconditional upper limits $m_{B}$ and $E_{B}$
of the mass and energy of the ejecta particle, respectively, 
which can be realized only if the emitted particle is massless.
The ratio of $E_{B}$ to $E_{1}$ 
the energy of incident critical particle 
takes a maximum $(2+\sqrt{3})(2-\sqrt{2})\simeq 2.186$,
for which the incident critical particle is massive and 
marginally bound. In general, 
the most energetic particle that escapes to 
infinity must be ejected inwardly on the production and 
subsequently bounces back outwardly at the turning point 
which is very close to the horizon
due to the angular momentum which is slightly above the critical value.
We have also determined the 
upper limit $\eta_{B}$ of the energy extraction efficiency
for the upper limit of ejecta energy
from the near-horizon collision 
with an arbitrarily high CM energy. 
$\eta_{B}$ is given by $(2+\sqrt{2}+\sqrt{6})/4\simeq 1.466$, which
can be realized for the collision of two marginally bound 
massive particles with optimized mass ratio.

We have next analyzed perfectly elastic collision,
Compton scattering, and pair annihilation. 
In all these cases, the energy of the emitted particle 
can be really greater than that of the injected critical particle.
We have also found that net positive energy 
extraction is not possible for perfectly elastic collision of 
equal masses, while it is possible for perfectly elastic collision 
with optimized mass ratio, 
Compton scattering, and pair annihilation.
In particular, the (inverse) Compton scattering of a subcritical 
photon by a critical massive particle is most efficient
among these three reactions as a physically realistic 
process of energy extraction. Although the present analysis 
is restricted in the equatorial plane, it is unlikely that 
the result would be drastically changed even if we allow 
non-equatorial reactions.

Jacobson and Sotiriou (2010)~\cite{Jacobson:2009zg} claim
that, for the collision of two particles of equal mass $m_{0}$,
the energy of the ejecta particle does not exceed $2m_{0}$ but 
drops to something just 
slightly above $m_{0}$ in the limit of infinite CM energy. 
The present result contradicts their claim.
As we have shown, the energy of the ejecta particle can be
$1.343m_{0}$ and $2.186m_{0}$ for perfectly elastic collision
of two equal masses and for pair annihilation, 
respectively, in the limit of infinite CM energy. 
The latter gives the unconditional energy upper limit 
for the collision of two equal masses and 
enables net positive energy extraction.
The disagreement of the claim in Ref.~\cite{Jacobson:2009zg} 
with the present result is probably due to 
the strong assumption adopted in Ref.~\cite{Jacobson:2009zg} 
that the four-momentum of the ejecta particle is 
parallel to that of the incident critical particle.
We think that this assumption is not valid in estimating the 
energy of the emitted particle.
See also Ref.~\cite{Grib:2010bs}.

The present result directly implies that when we consider 
gamma-ray emission from the pair annihilation of 
dark matter particles of mass $m$ near the rapidly 
rotating Kerr black hole, the spectrum due to the BSW collision 
continues up to 218.6 MeV $(m/100 \mbox{MeV})$ and is cut off there.
This is also the case for gamma-ray spectrum from the 
inverse Compton scattering by dark matter particles.

On the other hand, since the CM energy of particle collisions 
can be extremely high, high-energy reactions which are 
prohibited in low-energy collision may occur
and leave their signatures in relatively low-energy gamma-ray 
spectrum in general. In this context, it should be noted that
that Cannoni et al.~\cite{Cannoni:2012rv} 
discuss the possibility that colliding dark matter particles in the form of neutralinos 
may be gravitationally boosted near the supermassive black hole at the galactic center
so that they can have enough collision energy to annihilate into a stau pair
in some phenomenologically favored supersymmetric models. They also suggest the possibility 
that the signatures of the new channel of the reactions in gamma-ray spectrum might be discriminated by the Fermi-LAT 
satellite observation. They take into account the gravitational boost
with the relative velocity 
is $\sim 0.1-0.2$ light speed, which exists also for a non-rotating black hole. The CM energy can be $2\sqrt{5}m_0$ at maximum 
in the former effect, while it can be $\sim 19 m_0$ for $a/M=0.998$ in the 
latter effect~\cite{Harada:2010yv}, where $m_{0}$ is the mass of the dark matter particle. 
This strongly suggests the channel of the dark matter pair annihilation may also be opened through the BSW process near a rapidly rotating black hole in some supersymmetric models, although the detailed analysis with fully general relativistic treatment is yet to be done. 

The present analysis is restricted to a maximally rotating black 
hole, which is not expected to exist as an astrophysical object. 
It is interesting to study the upper limits
of particle emission from the high-energy collision near 
a non-maximally rotating black hole. 
However, we can naturally expect that the upper limits 
of the emission do not change so drastically even 
if $a/M$ is slightly below unity, although 
the maximum CM energy itself is sensitive to $a/M$.
This is because the present upper limits for 
the maximal rotation are finite and 
determined by the spacetime geometry near the horizon 
and the metric there can change only smoothly as $a/M$ 
increases to unity from below.

While the authors were finalizing the present paper, 
two papers~\cite{Zaslavskii:2012yp,Bejger:2012yb} 
appeared on the arXiv, in which the upper limits of the mass, 
energy, and energy 
extraction efficiency are studied. 
Although the present result is 
consistent with the result of Ref.~\cite{Bejger:2012yb},
not only does the present paper contain 
further new findings but also place the baseline for future
research on this subject because of its systematic and 
analytical approach.

\acknowledgments

T.H. thanks H.~Asada, T.~Igata, K.~T.~Inoue, M.~Kasai, M.~Kimura, Y.~Kojima, M.~Shimano, J.~Silk, R. Takahashi, and A.~Tomimatsu 
for helpful discussion. He would also thank the Astronomy Unit, Queen Mary, University of London for its hospitality.
T.H. was supported by Grant-in-Aid for Scientific Research from
the Ministry of Education, Culture, Sports, Science, and Technology of Japan [Young Scientists (B)
No. 21740190].  
The authors would like to thank the anonymous referee 
for helpful comments.

\appendix

\section{Collision with a circularly orbiting particle}
\label{sec:circular_orbit}

The expression for $m_{4}^{2}$ given by Eq.~(\ref{eq:m4_general}) 
contains a term which apparently diverges in the limit $E_{1}\to m_{1}/\sqrt{3}$.
To get a consistent approach, 
we here consider a massive particle circularly orbiting near the horizon 
because its energy approaches $m/\sqrt{3}$ and 
angular momentum asymptotically satisfies the critical condition
in the near-horizon limit as we will see below.
See Ref.~\cite{Pugliese:2011xn} for circular orbits in the 
extreme Kerr spacetime in more general context.
The energy and angular momentum of the circular orbit can be obtained by solving $V(r)=V'(r)=0$. Putting $r=M/(1-\epsilon)$ and 
solving $V(r)=V'(r)=0$ for $E$ and $\tilde{L}$ order by order, we obtain
\begin{equation}
E=\frac{m}{\sqrt{3}}\left(1+\frac{2}{3}\epsilon+\frac{1}{24}\epsilon^{2}\right)+O(\epsilon^{3})
\end{equation}
and 
\begin{equation}
\frac{\tilde{L}}{2E_{1}}=1+\frac{1}{4}\epsilon^{2}+\frac{1}{16}\epsilon^{3}+O(\epsilon^{4}).
\end{equation}
Assuming particle 1 belongs to this class, we find that $p_{1}^{r}=0$ by definition, while
\begin{eqnarray}
|p_{4}^{r}|&=& \left( 2E_{2}-L_{2} \right) + 
\left[ 2(L_{2}-E_{2})+2E_{3}\,(\delta_{(1)}-1)
+2E_{1}\right] \epsilon \nonumber \\
&&+\left[\frac{(2E_{2}-L_{2})}{2}-2(2\delta_{(1)}-\delta_{2})E_{3}
-\frac{[(m_{1}/\sqrt{3})+E_{2}-E_{3}]^{2}+m_{4}^{2}}{2(2E_{2}-L_{2})}+\frac{5}{6}\frac{m_{1}}{\sqrt{3}}
\right]\epsilon^{2}\nonumber \\
&&+O(\epsilon^{3}).
\end{eqnarray}
Then, Eq.~(\ref{eq:general_1st_order}) is not changed with $E_{1}=m_{1}/\sqrt{3}$, 
while Eq.~(\ref{eq:m4_general}) is changed to 
\begin{eqnarray}
m_{4}^{2}&=&(2E_{2}-\tilde{L}_{2})\left[\frac{5}{3\sqrt{3}}m_{1}
-4(2\delta_{(1)}-\delta_{(2)})E_{3}+2\sigma_{3}\frac{E_{3}^{2}[1-4(2\delta_{(1)}-\delta_{(2)})(1-\delta_{(1)})]}
{\sqrt{E_{3}^{2}(3-8\delta_{(1)}+4\delta_{(1)}^{2})-m_{3}^{2}}}\right] \nonumber \\
&&+(E_{2}^{2}+m_{2}^{2})-\left(\frac{m_{1}}{\sqrt{3}}+E_{2}-E_{3}
\right)^{2}.
\end{eqnarray}
The apparently divergent term in Eq.~(\ref{eq:m4_general}) 
in the limit $E_{1}\to m_{1}/\sqrt{3}$ is now replaced with a finite term. 
 
\section{Proof for the condition on the energy of particle 2
\label{sec:negative}}

First we calculate
\begin{equation}
\left[\frac{\lambda_{+}^{2}}{\sqrt{3\lambda_{+}^{2}-m_{3}^{2}}}
\right]^{2}
-\left[\frac{E_{1}^{2}}{\sqrt{3E_{1}^{2}-m_{1}^{2}}}\right]^{2}
=\frac{\lambda_{+}^{4}(3E_{1}^{2}-m_{1}^{2})-E_{1}^{4}(3\lambda_{+}^{2}-m_{3}^{2})}{(3\lambda_{+}^{2}-m_{3}^{2})(3E_{1}^{2}-m_{1}^{2})}.
\end{equation}
For $m_{3}=0$, the numerator can be written as follows:
\begin{eqnarray}
\lambda_{+}^{4}(3E_{1}^{2}-m_{1}^{2})-E_{1}^{4}(3\lambda_{+}^{2}-m_{3}^{2})&=&
[(3E_{1}^{2}-m_{1}^{2})\lambda_{+}^{2}-3E_{1}^{4}]\lambda_{+}^{2}+E_{1}^{4}m_{3}^{2} \nonumber \\
&=& E_{1}^{4}E_{B}^2 f(x),
\end{eqnarray}
where 
\begin{eqnarray}
f(x)=x^{2}\left(\frac{2-x}{2-\sqrt{3}}\right)^{2}-3, \quad x= \sqrt{3-\frac{m_{1}^{2}}{E_{1}^{2}}}.
\end{eqnarray}
For $E_{1}\ge m_{1}$, we find $\sqrt{2}\le x\le \sqrt{3}$.
$f(x)$ monotonically decreases in this domain 
and $f(\sqrt{3})=0$. Hence, $f(x)\ge 0$ for $\sqrt{2}\le x\le \sqrt{3}$. 
The condition (\ref{eq:e2_lower_bound}) follows from Eq.~(\ref{eq:m4_general_lambda+})
with $\delta_{(2)}\ge 0$, $\tilde{L}_{2}<2E_{2}$, $\lambda_{+}\ge m_{3}$, and $m_{4}^{2}\ge 0$. Q.E.D.

\end{document}